\documentclass[12pt]{article}

\usepackage{graphics}

\newcommand{\be}{\begin{equation}}
\newcommand{\ee}{\end{equation}}
\newcommand{\bea}{\begin{eqnarray}}
\newcommand{\eea}{\end{eqnarray}}

\begin{document}

\begin{titlepage}

\pagestyle{empty}

\setlength{\baselineskip}{18pt}

\vskip .2in

\begin{center}

{\large{\bf Relativistic Calculations for Incoherent Photoproduction of
$\eta$ Mesons}} \end{center}

\vskip .1in

\begin{center}
I. R. Blokland and H. S. Sherif 

{\it Department of Physics, University of Alberta}

{\it  Edmonton, Alberta, Canada T6G 2J1}

\vskip .1in

\end{center}

\centerline{ {\bf Abstract} }

\setlength{\baselineskip}{18pt}

\noindent

We develop a relativistic model for incoherent $\eta$-photoproduction on
nuclei. The elementary process is described using an effective
Lagrangian containing photons, nucleons, the $S_{11}(1535)$ and
$D_{13}(1520)$ nucleon resonances, and $\rho$, $\omega$, and $\eta$
mesons. The nucleon and $\eta$ wavefunctions are obtained from
relativistic wave equations. Final-state interactions of the outgoing
particles are included via optical potentials. The effects of these
interactions are found to be large and lead to reduced cross sections.The
incoherent cross sections for isovector transitions are much larger than 
those for isoscalar ones. The dominant contributions are those from the
$S_{11}$ and $D_{13}$ resonances. We find important interference 
effects between the contributions of these two resonances. 
We give some detailed calculations for the cross sections for incoherent
$\eta$-photoproduction on $^{12}$C. We find that the incoherent
cross section for a subset of states in the excitation energy region 
below $17\,\rm{MeV}$ are significantly larger than those
of the coherent process. These cross sections may thus
be accessible experimentally.  

\vskip .1in



\end{titlepage}

\setlength{\baselineskip}{18pt}



\section{Introduction}
\label{intro}

Over the past decade, $\eta$ meson photoproduction reactions have been
the subject of a number of investigations, both theoretical and
experimental. From a theoretical standpoint, $\eta$-photoproduction
provides a useful method to examine the properties of certain nucleon
resonances. In particular, since the $\eta$ meson is a spin- and
isospin-zero particle, its coupling to nucleons can lead to the
formation of only isospin-$\frac{1}{2}$ nucleon resonances. Using
nuclear targets, $\eta$-photoproduction can be used to investigate the
modification of hadron properties in the nuclear medium. These reactions
can also be used to study the final-state interactions of $\eta$ mesons
with nuclei, another topic of current interest. Reactions on nuclei can
complement and further test information obtained from studying
photoproduction reactions on free nucleons. 

In recent years, there have been several theoretical treatments of
coherent $\eta$-photoproduction reactions on nuclei. Fix and
Arenh{\"{o}}vel~\cite{fa} used an effective Lagrangian approach (ELA) to
examine the free $\eta$-photoproduction process, using only the
$S_{11}(1535)$ and $D_{13}(1520)$ resonances, as well as $t$-channel
vector meson exchange and nucleon pole terms. From these results, they
obtained cross sections for coherent $\eta$-photoproduction on $^{4}$He
and $^{12}$C in the near-threshold region. Peters {\it et
al.}~\cite{peters} used an ELA with a relativistic, non-local model to
study coherent $\eta$-photoproduction on spin-zero nuclei. They also
compared several optical potentials for the $\eta$ final-state
interactions. Piekarewicz {\it et al.}~\cite{psm1,psm2} used a
relativistic ELA to examine coherent $\eta$-photoproduction on $^{4}$He,
$^{12}$C, and $^{40}$Ca. Bennhold and Tanabe~\cite{benn} used a coupled
channel isobar model, in which the $(\gamma,\eta)$ reaction is related
to the $(\gamma,\pi)$, $(\pi,\eta)$, $(\pi,\pi)$, and $(\eta,\eta)$
reactions. They used the resulting elementary amplitude to study
coherent and incoherent photoproduction of $\eta$ mesons on nuclei. The
term incoherent photoproduction, as introduced by these authors, refers
to reactions leading to excited states of the final nucleus. We adopt
this same definition in the present work. 

The paper by Bennhold and Tanabe constitutes the only existing
theoretical treatment of incoherent $\eta$-photoproduction. The small
cross sections they obtained, relative to the dominant quasifree
$\eta$-photoproduction process, indicate that these processes are out of
reach for the current generation of experiments. Furthermore, nuclear
structure complications have curtailed theoretical interest in
incoherent reactions. To date, only inclusive measurements have been
made for $\eta$-photoproduction on complex nuclei~\cite{mami3}. In order
to understand the underlying mechanisms of the process better, these
measurements will need to be complemented by quasifree, coherent, and
incoherent measurements. 

In this paper we extend a previous relativistic model for quasifree
photoproduction of $\eta$ mesons on complex nuclei~\cite{hphs1,hphs2} to
the case of incoherent $\eta$-photoproduction. The main ingredients of
the present treatment are as follows. The effective Lagrangian of
Benmerrouche {\it et al.}~\cite{bmz} is used for the interactions
between fields. Contributions from the nucleon Born diagrams,
$t$-channel vector mesons, and the $S_{11}$ and $D_{13}$
nucleon resonances are included. The nuclear wavefunctions are solutions
of the Dirac equation with strong scalar and vector potentials in the
spirit of the relativistic mean field theory of
Walecka~\cite{walecka1,walecka2}. Calculations are carried out in the
plane wave approximation (PWA) and also in the distorted wave
approximation (DWA) in which the final-state interactions of the $\eta$
meson are taken into account.

In the following section, we outline the calculations for the amplitude
and observables of an incoherent $\eta$-photoproduction reaction. In
Sec.~\ref{parameters} we discuss the parameters that are used in the
effective Lagrangian. The results of the calculations are presented and
discussed in Sec.~\ref{results}. Conclusions are given in
Sec.~\ref{concs}.



\section{Reaction Model}
\label{model}

In an incoherent $\eta$-photoproduction reaction, a photon interacts
with a nucleus to produce an $\eta$ meson and raise the nucleus from its
ground state to an excited state. In the impulse approximation,
many-body contributions are neglected so that the production takes place
on a single nucleon. In this approximation, the transition amplitude for
the incoherent reaction $A(\gamma,\eta)A^{\ast}$ is closely related to
that of the elementary reaction $N(\gamma,\eta)N$.

The starting point in the present approach is a relativistic interaction
Lagrangian for a system of photons, mesons, nucleons, and nucleon
resonances from which one obtains the transition amplitude for the
$A(\gamma,\eta)A^{\ast}$ reaction. The amplitude is then used to
calculate the observables for the reaction.

In the photoproduction of $\eta$ mesons from complex nuclei, the
reaction takes place within the nuclear medium. The dynamics of the
nucleons within the nuclear matter are described by the relativistic
mean field Lagrangian of Walecka~\cite{walecka1,walecka2}. The $\eta$
meson is described by solutions of the Klein-Gordon equation. The
interactions of the fields involved in the reaction are described by the
interaction Lagrangian
\be
\label{l-int}
\mathcal{L}_{\rm{INT}} = 
     \mathcal{L}_{\eta NN} + \mathcal{L}_{\gamma NN} + 
     \mathcal{L}_{V \eta \gamma} + \mathcal{L}_{VNN} + 
     \mathcal{L}_{\eta NR} + \mathcal{L}_{\gamma NR}\; .
\ee
We use the effective Lagrangian of Benmerrouche {\it et al.}~\cite{bmz},
using pseudoscalar coupling for the $\eta NN$ vertex. The terms
in~(\ref{l-int}) can be written explicitly as:
\be
\label{l-enn}
\mathcal{L}_{\eta NN} = -i g_{\eta NN} \overline{\psi} \gamma_{5} \psi \eta \; ,
\ee
\be
\label{l-gnn}
\mathcal{L}_{\gamma NN} = -e \overline{\psi} \gamma_{\mu} A^{\mu} \psi - \frac{e\kappa_{p}}{4M} \overline{\psi} \sigma_{\mu \nu} F^{\mu \nu} \psi \; , 
\ee  
\be
\label{l-vnn}
\mathcal{L}_{VNN} = -g_{v} \overline{\psi} \gamma_{\mu} V^{\mu} \psi - \frac{g_{t}}{4M} \overline{\psi} \sigma_{\mu \nu} V^{\mu \nu} \psi \; , 
\ee
\be
\label{l-veg}
\mathcal{L}_{V\eta \gamma} = \frac{e\lambda_{v}}{4m_{\eta}} \epsilon_{\mu \nu \lambda \sigma} F^{\mu \nu} V^{\lambda \sigma} \eta \; .
\ee
At the $VNN$ vertex, we use a form factor of the type
\be
\label{ff}
F(t) = \frac{\Lambda^{2} - m_{V}^{2}}{\Lambda^{2} - t}
\ee
with $\Lambda^{2} = 1.2 \, \rm{GeV}^{2}$.
For the $S_{11}$ resonance,
\be
\label{l-ens}
\mathcal{L}_{\eta NR} = -i g_{\eta NS_{11}} \overline{\psi} R \eta + \rm{H.c.} \; ,
\ee
\be
\label{l-gns}
\mathcal{L}_{\gamma NR} = - \frac{e \kappa_{S_{11}}}{2(M + M_{S_{11}})} \overline{R} \gamma_{5} \sigma_{\mu \nu} F^{\mu \nu} \psi + \rm{H.c.} \; .
\ee
For the $D_{13}$ resonance,
\be
\label{l-end}
\mathcal{L}_{\eta NR} = \frac{f_{\eta ND_{13}}}{m_{\eta}} \; \overline{R}^{\mu} \theta_{\mu \nu}(Z) \gamma_{5} \psi \partial^{\nu}\eta + \rm{H.c.} \; ,
\ee
\be
\label{l-gnd1}
\mathcal{L}_{\gamma NR}^{(1)} = \frac{ie\kappa^{(1)}_{D_{13}}}{2M} \; \overline{R}^{\mu} \theta_{\mu \nu}(Y) \gamma_{\lambda} \psi F^{\nu \lambda} + \rm{H.c.} \; ,
\ee
\be
\label{l-gnd2}
\mathcal{L}_{\gamma NR}^{(2)} = \frac{e\kappa^{(2)}_{D_{13}}}{4M^{2}} \; \overline{R}^{\mu} \theta_{\mu \nu}(X) \partial_{\lambda} \psi F^{\nu \lambda} + \rm{H.c.} \; ,
\ee
where the tensors $V^{\mu \nu}$ and $\theta_{\mu \nu}(V)$ are defined by
\be
\label{v}
V_{\mu \nu} = \partial_{\mu}V_{\nu} - \partial_{\nu}V_{\mu} \; ,
\ee
\be
\label{theta}
\theta_{\mu \nu}(V) = g_{\mu \nu} + \left[ - \frac{1}{2} \; (1 + 4V) + V \right] \gamma_{\mu}  \gamma_{\nu} \; ,
\ee
for $V=X,Y,Z$.  We take $X,Y,Z=-1/2$ so that the $D_{13}$ terms of $\mathcal{L}$ agree with Peters {\it et al.}~\cite{peters}

At tree level, the $S$ matrix for the $A(\gamma,\eta)A^{\ast}$ reaction is
\be
\label{sfi}
S_{fi} = - \frac{1}{2} \int \left< f \left| T \left[ \mathcal{L}_{\rm{INT}}(x) \mathcal{L}_{\rm{INT}}(y) \right] \right| \! i \right> d^{4}x \; d^{4}y \; .
\ee
Following the procedure in~\cite{hphs1}, we obtain
\bea
\label{sfi2}
S_{fi} & = & \sum_{(j)} \frac{e}{(2\pi)^{3}} \sqrt{\frac{1}{2E_{\eta} \; 2E_{\gamma}}} \sum_{J_{C} J_{B} M_{B}} \sum_{J_{B^{\prime}} M_{B^{\prime}}} (J_{C},J_{B}; M_{C},M_{B} | J_{i},M_{i}) \nonumber \\
& & \times (J_{C},J_{B^{\prime}}; M_{C},M_{B^{\prime}} | J_{f},M_{f}) \left[ \mathcal{S}_{J_{i}J_{C}}(J_{B}) \right] ^{1/2} \left[ \mathcal{S}_{J_{f}J_{C}}(J_{B^{\prime}}) \right] ^{1/2} \nonumber \\ & & \times \int d^{4}x \Psi_{J_{B^{\prime}} M_{B^{\prime}}}^{\dagger} (x) \; \Gamma_{(j)} \; \Psi_{J_{B} M_{B}} (x) \; \Phi_{\eta}^{\ast}(x) \; e^{-i k_{\gamma} \cdot x} \; .
\eea
In the above equation, $E_{\gamma}$ and $E_{\eta}$ are the energies of
the photon and $\eta$ meson. The struck nucleon has angular momentum
quantum numbers $J_{B}$ and $M_{B}$ before the interaction and
$J_{B^{\prime}}$ and $M_{B^{\prime}}$ after the interaction. $J_{C}$ and
$M_{C}$ denote the angular momentum quantum numbers of the nuclear core,
which is defined to comprise the remainder of the nucleus. The nucleus
has angular momentum quantum numbers $J_{i}$ and $M_{i}$ before the
interaction and $J_{f}$ and $M_{f}$ after the interaction.
$(J_{C},J_{B}; M_{C},M_{B} | J_{i},M_{i})$ is a Clebsch-Gordan
coefficient and $\mathcal{S}$ denotes a spectroscopic factor.
$\Psi_{J_{B} M_{B}}(x)$ and $\Psi_{J_{B^{\prime}} M_{B^{\prime}}}(x)$
represent the wavefunctions of the bound nucleon, before and after the
interaction, and $\Phi_{\eta}(x)$ is the $\eta$ meson wavefunction.
Finally, $\Gamma_{(j)}$ is a $4\times 4$ matrix operator which contains
the details of the interaction relevant to a particular reaction
channel, labeled by $(j)$. The explicit forms for
$\Gamma_{\rm{proton}}$, $\Gamma_{S_{11}}$, $\Gamma_{D_{13}}$, and
$\Gamma_{V}$ are given in~\cite{hphs1}. 

We carry out two types of calculations depending on whether or not the
final-state interactions of the $\eta$ are taken into account. In the
PWA, we neglect final-state interactions so the $\eta$ meson
wavefunction takes the form
\be
\label{eta-pwa}
\Phi_{\eta}(x) = e^{-ik_{\eta}\cdot x} .
\ee 
In the DWA, the $\eta$ meson wavefunction is distorted through the use
of an optical potential in the Klein-Gordon equation. In our DWA
calculations, we will use two different optical potentials. The first
optical potential, which we will denote DW1, was introduced by Lee {\it
et al.}~\cite{lee} using the $\eta N$ scattering amplitude found by
Bennhold and Tanabe~\cite{benn}. The second optical potential, which we
will label DW3, was introduced by Peters {\it et al.}~\cite{peters}
using the results of Effenberger and Sibirtsev~\cite{eff}.

In order to write expressions for the observables of the reaction, we
will find it useful to define a function $\mathcal{Z}$ as
\be
\label{z}
\mathcal{Z}_{(j)} = \int d^{3}x \psi_{J_{B^{\prime}} M_{B^{\prime}}}^{\dagger} (\vec{x}) \; \Gamma_{(j)} \; \psi_{J_{B} M_{B}} (\vec{x}) \; \varphi_{\eta}^{\ast}(\vec{x}) \; e^{i \vec{k}_{\gamma} \cdot \vec{x}}\; ,
\ee
where $\psi$ and $\varphi$ are the spatial parts of the particle
wavefunctions. Equation (\ref{sfi2}) can then be written as
\bea
\label{sfi3}
S_{fi} & = & \sum_{(j)} \frac{e}{(2\pi)^{2}} \sqrt{\frac{1}{2E_{\eta} \; 2E_{\gamma}}} \; \delta (E_{B^{\prime}} + E_{\eta} - E_{B} - E_{\gamma}) \nonumber \\ & & \times \; \sum_{J_{C} J_{B} M_{B}} \sum_{J_{B^{\prime}} M_{B^{\prime}}} (J_{C},J_{B}; M_{C},M_{B} | J_{i},M_{i}) \nonumber \\
& & \times (J_{C},J_{B^{\prime}}; M_{C},M_{B^{\prime}} | J_{f},M_{f}) \left[ \mathcal{S}_{J_{i}J_{C}}(J_{B}) \right] ^{1/2} \nonumber \\ & & \times \left[ \mathcal{S}_{J_{f}J_{C}}(J_{B^{\prime}}) \right] ^{1/2} \mathcal{Z}_{(j)} \; .
\eea
The differential cross section is then related to $\mathcal{Z}_{(j)}$ by
\be
\label{dc}
\frac{d\sigma}{d\Omega_{\eta}} = \frac{\alpha}{8\pi} \; 
\frac{(2J_{f} + 1)}{R} \;  
\frac{p_{\eta}}{E_{\gamma}} \sum_{(j)} \sum_{J_{B}, M_{B}} \sum_{J_{B^{\prime}}, M_{B^{\prime}}} \sum_{\xi} \frac{\mathcal{S}_{J_{i}J_{C}}(J_{B})}{(2J_{B} + 1)}
\frac{\mathcal{S}_{J_{f}J_{C}}(J_{B^{\prime}})}{(2J_{B^{\prime}} + 1)}
\left| \mathcal{Z}_{(j)} \right| ^{2} \; .
\ee
Note that $\mathcal{Z}$ depends on~$M_{B}, M_{B^{\prime}}$, and the
photon polarization~$\xi$. The recoil factor $R$ is given by
\be
\label{recoil}
R = 1 + \frac{E_{\eta}}{E_{R}} 
    \left( 1 - \frac{p_{\gamma}}{p_{\eta}} \; \cos \theta_{\eta} \right) \; ,
\ee
where $p_{\gamma}$ and $p_{\eta}$ are the momenta of the the photon and the
$\eta$-meson, respectively.  The photon asymmetry for linearly polarized incident photons is
\be
\label{sigma}
\Sigma = \frac{d\sigma_{\perp} - d\sigma_{\parallel}}{d\sigma_{\perp} + d\sigma_{\parallel}} \; ,
\ee
where $d\sigma_{\perp}$ and $d\sigma_{\parallel}$ are the differential
cross sections for specified polarizations of the incident photon,
namely, perpendicular and parallel to the plane of the reaction.



\section{Parameters}
\label{parameters}

The effective Lagrangian in Sec.~\ref{model} contains a number of
parameters, such as coupling constants and anomalous magnetic moments,
which must be inferred from the experimental results of other reactions.
In particular, experimental studies of the elementary
$\eta$-photoproduction reaction $p(\gamma,\eta)p$ allow us to constrain
our parameter set.

The parameters relating to the $S_{11}$ and $D_{13}$
resonances are the most crucial inputs for our effective Lagrangian. By
considering the decays of these resonances through the $\gamma p$ and
$\eta p$ channels, we can relate the parameters in
equations~(\ref{l-ens}) through~(\ref{l-gnd2}) to more conventional
resonance parameters~\cite{bmz}. For example, for the $S_{11}$
resonance,
\be
\label{ek-s11}
\left| e\kappa_{S_{11}} \right| = \sqrt{\frac{2M(M_{S_{11}} + M)}{(M_{S_{11}} - M)}} \; \left| A^{p}_{1/2} \right| \; ,
\ee
\be
\label{g-s11}
\left| g_{\eta N S_{11}} \right| = \left( \frac{4\pi M_{S_{11}}}{p_{\eta} (E_{N} + M)} \; \Gamma_{S_{11} \rightarrow \eta N} \right) ^{1/2} \; ,
\ee
where $A^{p}_{1/2}$ is a helicity amplitude, and $p_{\eta}$ and $E_{N}$
are the momentum of the $\eta$ and the energy of the nucleon,
respectively, in the center of momentum frame for the decay
$S_{11}\rightarrow \eta N$. Similar expressions can be written
for $f_{\eta ND_{13}}\kappa_{D_{13}}^{(1)}$ and $f_{\eta
ND_{13}}\kappa_{D_{13}}^{(2)}$. While these expressions specify the
magnitudes of the resonance parameters, they do not provide us with any
information about their phases.  
 
In the present analysis we shall compare three slightly different sets of coupling 
parameters for the effective Lagrangians involving the resonances. Fix and 
Arenh{\"{o}}vel~\cite{fa} obtained a set of parameters that gave a good description 
of the elementary cross sections measured at Mainz~\cite{mami1}. We were able to reproduce 
their results using a certain set of phases for the extracted coupling parameters. As a 
further test of these parameters, we compared our predictions for
the photon asymmetry of the elementary process to the results obtained
by a recent experiment at the ESRF~\cite{esrf}. Figure~\ref{figure1}
shows our prediction, along with the experimental results, for the
photon asymmetry when $E_{\gamma}=740\,\rm{MeV}$. The parameters are 
listed as set 1 in Table~\ref{res-params}. Note that this set is given only for protons;
the handling of neutron cross sections in this particular case is explained in the 
following section. Peters {\it et
al.}~\cite{peters} have also used the Fix and Arenh{\"{o}}vel parameters 
to extract the 
coupling parameters for both protons and neutrons. The proton parameters are only 
slightly different from those mentioned above and the neutron parameters
are based on the 1996 listings of the Particle Data Group (these are also 
the same in the 2000 listings). 
These coupling parameters are listed as set 2 in Table~\ref{res-params}. 
The proton parameters in 
Set 3 in Table~\ref{res-params} are based on the recent analysis of Tiator {\it et
al.}~\cite{tiator} of several observables for the photoproduction process 
on proton targets.
This analysis produced new parameters for the $D_{13}$. In particular these
authors find a much smaller branching ratio than that used by Fix and 
Arenh{\"{o}}vel.  The rest of the coupling parameters are the same as for set 2. We shall 
compare the cross sections calculated using these three sets of parameters.


\section{Results and Discussion}
\label{results}

In order to put our final results into proper perspective we need to  
 discuss two relevant issues. The cross sections for the incoherent 
process depend strongly on the isospin of the excited state. For 
isospin-zero targets we calculate these cross sections for 
both $\,T_{f}=0\,$ and $\,T_{f}=1\,$ 
nuclear excited states. We explain here how these calculations are 
done for the three sets of coupling parameters discussed above. Set 1 lists only the 
coupling parameters for 
protons (see Table~\ref{res-params}). The calculations of the isospin dependent cross 
sections in this case make use of the experimental results from Mainz 
for $\eta$-photopoduction on the proton~\cite{mami1} and the 
deuteron~\cite{mami2}. An analysis 
by these authors, based on the dominance of the $S_{11}$ resonance, 
established
a neutron to proton cross section ratio. From this ratio the amplitude 
for the elementary process can be decomposed into isovector and isoscalar 
components. Specifically, based on the analysis by Krusche {\it et
al.}~\cite{mami2}, we 
take the ratio of the neutron to proton amplitudes to be $-0.80$. In our 
calculations we use the parameters of set 1 to obtain the contributions 
from the target protons to the incoherent amplitudes and use the above 
ratio to calculate the neutron contributions. These amplitudes can then 
be combined to yield the $\,T=0\,$ or
 $\,T=1\,$  amplitudes.  This procedure is 
used only with set 1; sets 2 and 3 have separate coupling parameters for 
protons and neutron and the amplitudes are calculated independently.

The other issue that also has some bearing on the calculated cross 
sections is the choice 
of the initial and final momenta of the participating nucleon. We
shall make a comparison between two choices. 
One is the ``frozen nucleon" approximation in which each nucleon moves 
as part of the target but without allowance for the Fermi motion. The 
second, and possibly more appropriate, choice for the nucleon momenta 
allows for Fermi motion using an approximate model employed by other 
authors~\cite{tiator2,mach,kam}. 
We use this model in the following form: The initial nucleon momentum in 
the lab frame is taken as: 
$\mbox{\boldmath{$p$}}_{i} = \frac{A-1}{2A}\left(\mbox{\boldmath{$p$}}_{\eta} 
                             - \mbox{\boldmath{$p$}}_{\gamma}\right)$. This choice is the 
same as the effective momentum used in refs~\cite{fa,benn,try}. 
The momentum 
of the nucleon in the final state is obtained by applying momentum 
conservation in the elementary production process.

Calculations were carried out as outlined in Sec.~\ref{model} for
incoherent $\eta$-photoproduction cross sections on $^{12}$C, $^{16}$O,
and $^{40}$Ca. Qualitatively, our calculations 
show similar results for all three nuclei. Calculations on $^{16}$O and 
$^{40}$Ca
do not produce any significant
features beyond what we observe on  $^{12}$C. We therefore limit our
discussion to the latter nucleus. We will present the results of the 
$^{12}$C calculations,
in which we have used the $1p$-shell spectroscopic factors of Cohen and
Kurath~\cite{ck}. We have selected four excited states of $^{12}$C that
are well described by a $1p_{3/2}^{-1}-1p_{1/2}$ configuration:
$(2^{+}0;4.44)$, $(2^{+}1;16.11)$, $(1^{+}0;12.71)$, and
$(1^{+}1;15.11)$. It is found that transitions to the
$\,T=1\,$ states are much 
stronger than those to $\,T=0\,$ states. For this reason we will present a 
number of comparisons below involving calculations for only
the $(2^{+}1;16.11)$ state. Additional calculations involving all of the 
four states will also be presented.
  
We now discuss the results of our calculations for the
$^{12}$C$(\gamma,\eta)^{12}$C$^{\ast}(2^{+}1;16.11)$ reaction.
We begin by looking at the dependence of the cross sections on the 
choice of nucleon momenta and the 
sensitivity to the three coupling parameter choices discussed 
above. Using parameter set 1, we show in Figure~\ref{figure2} a 
comparison between 
the two choices of nucleon momenta. The calculations, carried out
using the PWA, are presented for the 
differential cross section in the laboratory frame at 
$E_{\gamma}=766\,\rm{MeV}$.
The frozen nucleon approximation leads 
to larger cross sections in the forward direction than those for which the 
Fermi motion is taken into account in the approximate manner discussed above. The shapes 
of the two angular distributions are very similar.  Of particular interest, though, is that the relative contributions of the $S_{11}$ and $D_{13}$ resonances are strongly influenced by the choice of nucleon momenta, even though the total cross sections are reasonably stable. 
Because the allowance for Fermi motion 
is the more realistic choice, all subsequent calculations in 
this paper are carried out with this choice. 

In Figure~\ref{figure3} we compare the cross sections for the other 
two sets of coupling parameters (sets 2 and 3, Table~\ref{res-params}). 
The separate 
contributions of the $S_{11}$ and $D_{13}$ are also shown. 
Note that in the present model the background terms do
not contribute to isovector transitions, as the coupling parameters used for these terms~\cite{peters} are the same for protons and neutrons. The cross sections (dotted curve) 
due to the $S_{11}$ are the same for both sets since the two sets 
have identical parameters for this resonance. 
The contributions from the $D_{13}$ are much 
smaller, with those from set 2 being almost double those from set 3. This
reflects the small branching ratios extracted by the analysis of 
Tiator {\it et al.}~\cite{tiator}. This effect occurs despite the observation that the couplings $f_{\eta ND_{13}}\kappa_{D_{13}}^{(1)}$ and $f_{\eta ND_{13}}\kappa_{D_{13}}^{(2)}$ are larger for the set 3 parameters than for set 2.  We have found that these two parameters give rise to amplitudes of opposite sign and that the cancellation is stronger for the set 3 parameters.
One important feature evident in
Figure~\ref{figure3} is the strong constructive 
interference between the $S_{11}$ and $D_{13}$ contributions in this energy
region. Even though the $D_{13}$ cross section is by itself small, the 
combined cross section is strongly enhanced 
(see below for a discussion of the energy dependence of this 
interference effect).  The cross section curves in Figure~\ref{figure3} 
should also be compared with the thick solid curve in 
Figure~\ref{figure2} which gives the corresponding result for parameter 
set 1.  We see that the 
latter cross section falls in between those for sets 2 and 3. This gives
a measure of the adequacy of the assumption used together with set 1, namely 
that the isospin dependence of the total incoherent amplitude is the same 
as the $S_{11}$ amplitude.

From this point on all calculations will use parameter set 3 and 
allow for Fermi motion. Based on the comparisons presented above, 
this choice, in addition to being more realistic, should provide 
conservative estimates of the incoherent cross sections. 

It is now instructive to look at the energy dependence of the 
interference
between the $S_{11}$ and $D_{13}$ contributions. For the same state as above we 
show the results in Figure~\ref{figure4}, using parameter set 3 and 
again using plane waves for the outgoing eta particles. 
We see from the figure that the interference is destructive for 
energies near the threshold region. At photon energies above about 675 MeV
the interference  
pattern is constructive and the influence of the $D_{13}$ 
is somewhat enhanced. 
Calculations with sets 1 and 2 show similar behavior with slight changes 
in the energy at which transition from destructive to constructive 
interference takes place. For all parameter sets the total cross sections 
peak near $E_{\gamma}=750\,\rm{MeV}$. 
 
The cross sections in the above comparisons are calculated in the plane 
wave approximation in which the final-state interactions of the $\eta$ mesons
with the residual nucleus are 
neglected. In Figure~\ref{figure5} we study the effect of these 
final-state interactions. The calculations show the energy 
dependence of the total cross section for the same reaction on $^{12}$C. 
The two types of optical potentials referred to earlier are used to 
calculate the distorted waves of the $\eta$ mesons. 
The calculations show that the final-state 
interactions are substantial, leading to suppression of the cross sections 
particularly for photon energies near the peak region. For both 
distorting potentials
we observe a slight shift of the cross section peak towards
higher energies. At higher energies, the effects of the two optical
potentials begin to diverge. The DW1 potential
weakens at these energies whereas the DW3 potential retains its
strength and hence continues to suppress the cross section at these 
energies. The figure also shows the separate contributions from the 
$S_{11}$ and $D_{13}$ resonances for the DW1 calculations. We see 
that, as was the case 
for the PW calculations in Figure~\ref{figure4}, the 
interference between these two contributions changes from destructive 
near threshold to constructive at higher energies.

In Figure~\ref{figure6}, we compare the results of our calculations for
the differential cross sections of the
$^{12}$C$(\gamma,\eta)^{12}$C$^{\ast}(2^{+}0;4.44)$ and
$^{12}$C$(\gamma,\eta)^{12}$C$^{\ast}(2^{+}1;16.11)$ reactions at
$E_{\gamma}=650\,\rm{MeV}$ to the corresponding calculations by Bennhold
and Tanabe~\cite{benn}. Our calculations use optical potential DW1
and the coupling parameters of set 3. Based on the treatment of the bound 
nucleons, we
shall refer to our calculations as relativistic and Bennhold and
Tanabe's as nonrelativistic. There is good qualitative agreement between
the two sets of calculations. The relativistic calculations indicate a
larger cross section for the isovector transition than the
nonrelativistic calculations. The reverse holds for the 
isoscalar transition.
Furthermore, the suppression of the cross
sections due to final-state interactions is more significant in the
relativistic calculations. We note however that the two approaches
differ in many respects, for example, in the elementary input and the
form of the transition matrix elements.

From an experimental standpoint, it would be exceedingly difficult to
resolve a particular nuclear excited state left in the wake of an
incoherent $\eta$-photoproduction reaction. It might be possible,
however, to determine indirectly the excitation energy of the nucleus
with sufficient precision to exclude coherent and quasifree reactions.
To this effect, we have calculated the summed total incoherent
$\eta$-photoproduction cross section on $^{12}$C for $E_{x}$ in the
range of $4\,\rm{MeV}$ to $17\,\rm{MeV}$ by including all four excited
states mentioned at the beginning of this section. We have restricted 
the calculations to
excitation energies where we could treat the single-particle states as
bound states. 
Figure~\ref{figure7} shows this total cross section as a function of the
energy of the incident photon. The curves are DWA calculations using the
two optical potentials referred to earlier. This total cross section is
on the order of $100\,\rm{nb}$ or more for $750\,\rm{MeV}$ photons, 
which is
about an order of magnitude larger than several theoretical estimates of
the coherent cross section\cite{fa,peters} and only about two orders of
magnitude below the measured inclusive cross section for this target
nucleus~\cite{mami3}.



\section{Conclusions}
\label{concs}

In this paper we have developed a relativistic model for the incoherent
photoproduction of $\eta$ mesons. The ingredients of the model are that
(i) the nucleon wavefunctions are solutions of the Dirac equation with
appropriate scalar and vector potentials consistent with the
relativistic mean field approach, (ii) the $\eta$ meson is described by
solutions of the Klein-Gordon equation with appropriate optical
potentials, and (iii) the interactions between the fields are introduced
through a covariant effective Lagrangian. Contributions from the
$S_{11}(1535)$ and $D_{13}(1520)$ nucleon resonances, $t$-channel vector
mesons, and nucleon Born diagrams are included. The contributions by the latter two diagrams
to the incoherent cross sections are very small. 

Unlike coherent photoproduction on isoscalar targets,
the $S_{11}$ nucleon resonance provides the largest
contribution to the incoherent cross section. One of the interesting
results of the present work however, is the role of the $D_{13}$
resonance. Although its cross section is evidently smaller than that of 
the $S_{11}$, the $D_{13}$ has a significant effect on the cross
section through its interference with the $S_{11}$ contribution.  
We have established that this is independent of the set of coupling
parameters used. The interference pattern is destructive in the
threshold region, but becomes constructive at higher photon energies. The 
incoherent cross section is found to be much larger for 
isovector transitions than for isoscalar ones. This is consistent with
the results of the non-relativistic calculations of Bennhold and Tanabe.
We find however that the cross sections for the $\,T =1\,$ states are 
somewhat larger 
in our calculations, while the opposite is true for the $\,T =0\,$ 
transitions.  Furthermore
comparisons show that the suppression of the cross section due to
final-state inteactions is somewhat stronger in the relativistic approach.

Although the cross sections for an incoherent $\eta$-photoproduction
reaction to $\,T =0\,$ excited states are quite
small, those for $\,T =1\,$ excited states are found to be relatively large.
Our calculations show that the summed total incoherent
cross sections to a set of states in $^{12}$C, with excitation energies in the range $4-17\,\rm{MeV}$, are in excess of 100 nb. These are 
sufficiently large to be potentially observable. Such measurements would be 
valuable in  clarifying the seemingly enhanced role played by the $D_{13}$ 
in incoherent photoproduction.

\vskip .1in


\noindent{ {\bf Acknowledgements} }

\noindent 
We are grateful to M. Hedayati-Poor, J. I. Johansson, and R. L. Workman
for helpful communications concerning their work. This work was
supported in part by the Natural Sciences and Engineering Research
Council of Canada.

\newpage

\begin{center}
\textbf{Tables}
\end{center}

\begin{table}[ht]
\begin{center}
\begin{tabular}{l c c c}
& \textbf{Set 1} & \textbf{Set 2} & \textbf{Set 3} \\
& $p$ & $p/n$ & $p/n$ \\
\hline
$g_{\eta N S_{11}}$ & 2.0846 & 2.1/2.1 & 2.1/2.1 \\
$\kappa_{S_{11}}$ & -0.958 & -0.962/+0.817 & -0.962/+0.817 \\
$f_{\eta ND_{13}}\kappa_{D_{13}}^{(1)}$ & 37.75 & 36.9/-6.56 & 42.47/-6.56 \\
$f_{\eta ND_{13}}\kappa_{D_{13}}^{(2)}$ & 40.0 & 38.9/4.46 & 51.24/4.46 \\
\hline
\end{tabular}
\caption{$S_{11}(1535)$ and $D_{13}(1520)$ resonance parameters used in
our calculations.}
\label{res-params}
\end{center}
\end{table}


\newpage

\begin{center}
\textbf{Figure Captions}
\end{center}

\vskip .1in

\begin{figure}[!ht]
\caption{The photon asymmetry $\Sigma$, as a function of the center of
mass angle $\Theta_{\eta}$, for the elementary $\eta$-photoproduction
reaction $p(\gamma,\eta)p$ with $E_{\gamma}=740\,\rm{MeV}$. The present
calculations using parameter set 1 are represented by the solid line. 
The data points are from the
ESRF experiment~\cite{esrf}.}
\label{figure1}
\end{figure}

\begin{figure}[!ht]
\caption{The incoherent differential
cross section, as a function of the laboratory angle $\theta_{\eta}$,
for the $^{12}$C$(\gamma,\eta)^{12}$C$^{\ast}(2^{+}1;16.11)$ reaction at
$E_{\gamma}=766\,\rm{MeV}$. The curves are PWA
calculations using parameter set 1 according to the prescription explained in the text. The solid, dashed, and dash-dotted curves show the total, $S_{11}$, and $D_{13}$ cross sections, respectively.  The thin curves represent calculations in the frozen approximation for the struck 
nucleon kinematics and the thick curves show the results when Fermi motion 
is taken into account in the approximate manner discussed in the text.}
\label{figure2}
\end{figure}

\begin{figure}[!ht]
\caption{Same reaction as Figure~\ref{figure2} except that the 
calculations are for 
parameter sets 2 and 3. The thin solid curve (set 2) and thick solid curve
(set 3) are full PWA cross sections. The dotted curve shows the $S_{11} $
cross sections (identical for both sets). The dot-dashed
(dashed) curves show the $D_{13}$ cross sections for set 2 (set 3).} 
\label{figure3}
\end{figure}

\begin{figure}[!ht]
\caption{Total incoherent cross section for the 
$^{12}$C$(\gamma,\eta)^{12}$C$^{\ast}(2^{+}1;16.11)$ reaction as a 
function of the photon energy. The long- (short-) dashed curves show the 
separate cross sections of the $S_{11}$ ($D_{13}$) diagrams. 
The calculations are carried out using the PWA and parameter set~3.} 
\label{figure4}
\end{figure}

\begin{figure}[!ht]
\caption{Total incoherent cross section for the 
$^{12}$C$(\gamma,\eta)^{12}$C$^{\ast}(2^{+}1;16.11)$ reaction as a 
function of the photon energy.  The solid curve shows the PWA calculations.
The long-dashed (short-dashed) curves
are 
distorted wave calculations using optical potential DW1 (DW3).
The dotted (dash-dot) curves show the individual contributions of the 
$S_{11}$ ($D_{13}$) resonances.}
\label{figure5}
\end{figure} 

\begin{figure}[!ht]
\caption{Differential cross sections for the
$^{12}$C$(\gamma,\eta)^{12}$C$^{\ast}(2^{+}0;4.44)$ and
$^{12}$C$(\gamma,\eta)^{12}$C$^{\ast}(2^{+}1;16.11)$ reactions, denoted
by $T=0$ and $T=1$, respectively, with $E_{\gamma}=650\,\rm{MeV}$. The
upper graph, labeled {\it Relativistic}, shows the results of the
present calculations using parameter set 3. The lower graph, labeled {\it Nonrelativistic}, shows the
corresponding calculations by Bennhold and Tanabe~\cite{benn}. The
dashed curves indicate PWA calculations and the solid curves denote DWA
calculations using the DW1 optical potential.} 
\label{figure6}
\end{figure}

\begin{figure}[!ht]
\caption{Total cross sections for the incoherent $\eta$-photoproduction
reaction $^{12}$C$(\gamma,\eta)^{12}$C$^{\ast}$. Contributions from the
$^{12}$C$(\gamma,\eta)^{12}$C$^{\ast}(2^{+}0;4.44)$,
$^{12}$C$(\gamma,\eta)^{12}$C$^{\ast}(2^{+}1;16.11)$,
$^{12}$C$(\gamma,\eta)^{12}$C$^{\ast}(1^{+}0;12.71)$, and
$^{12}$C$(\gamma,\eta)^{12}$C$^{\ast}(1^{+}1;15.11)$ reactions are
included. The curves are distorted wave calculations using DW1 
(solid curve)
and DW3 (dashed curve) potentials.}
\label{figure7}
\end{figure}

\newpage
\begin{center}
\resizebox{\textwidth}{!}{\includegraphics{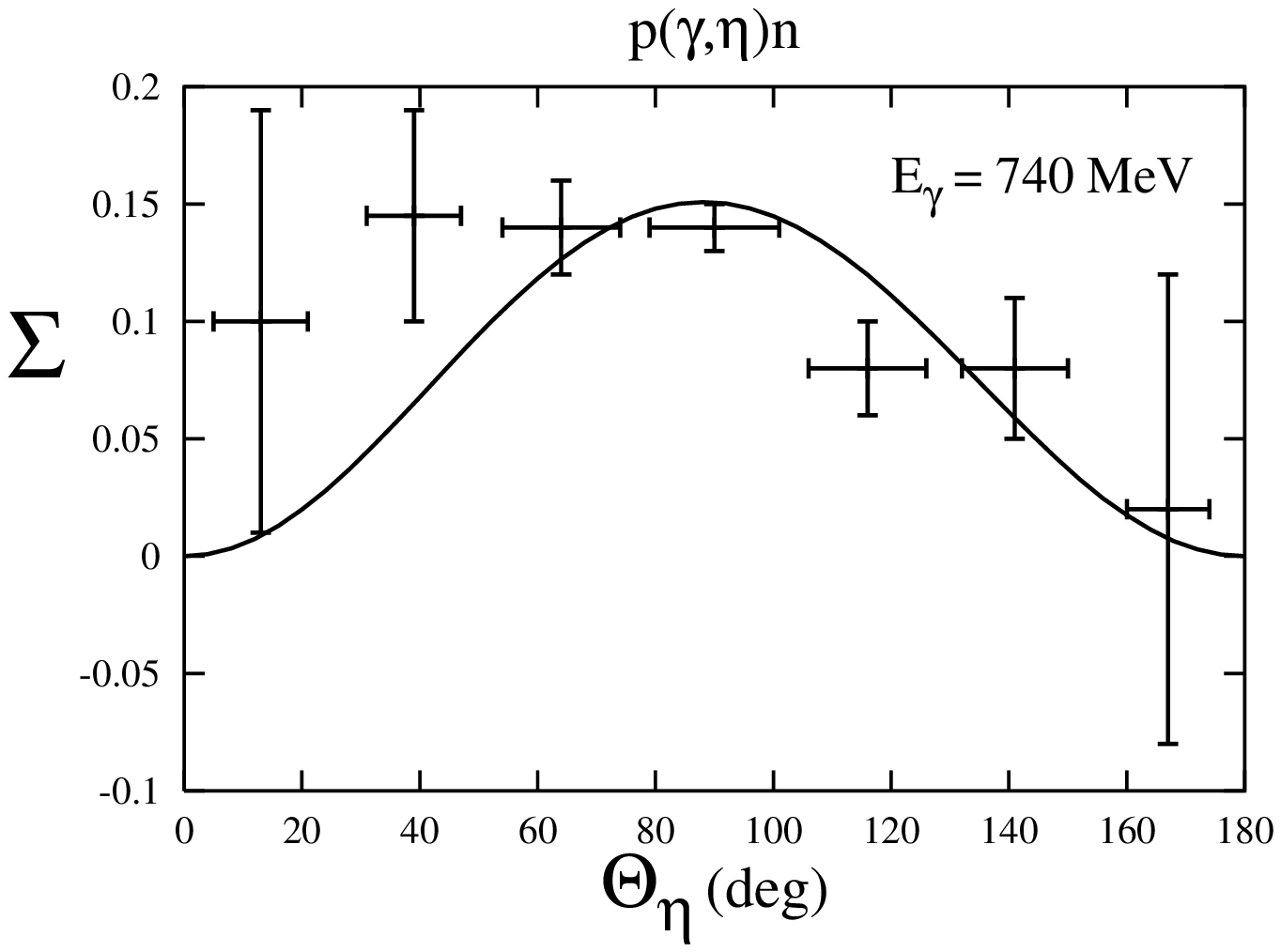}}
\end{center}

\newpage
\begin{center}
\resizebox{\textwidth}{!}{\includegraphics{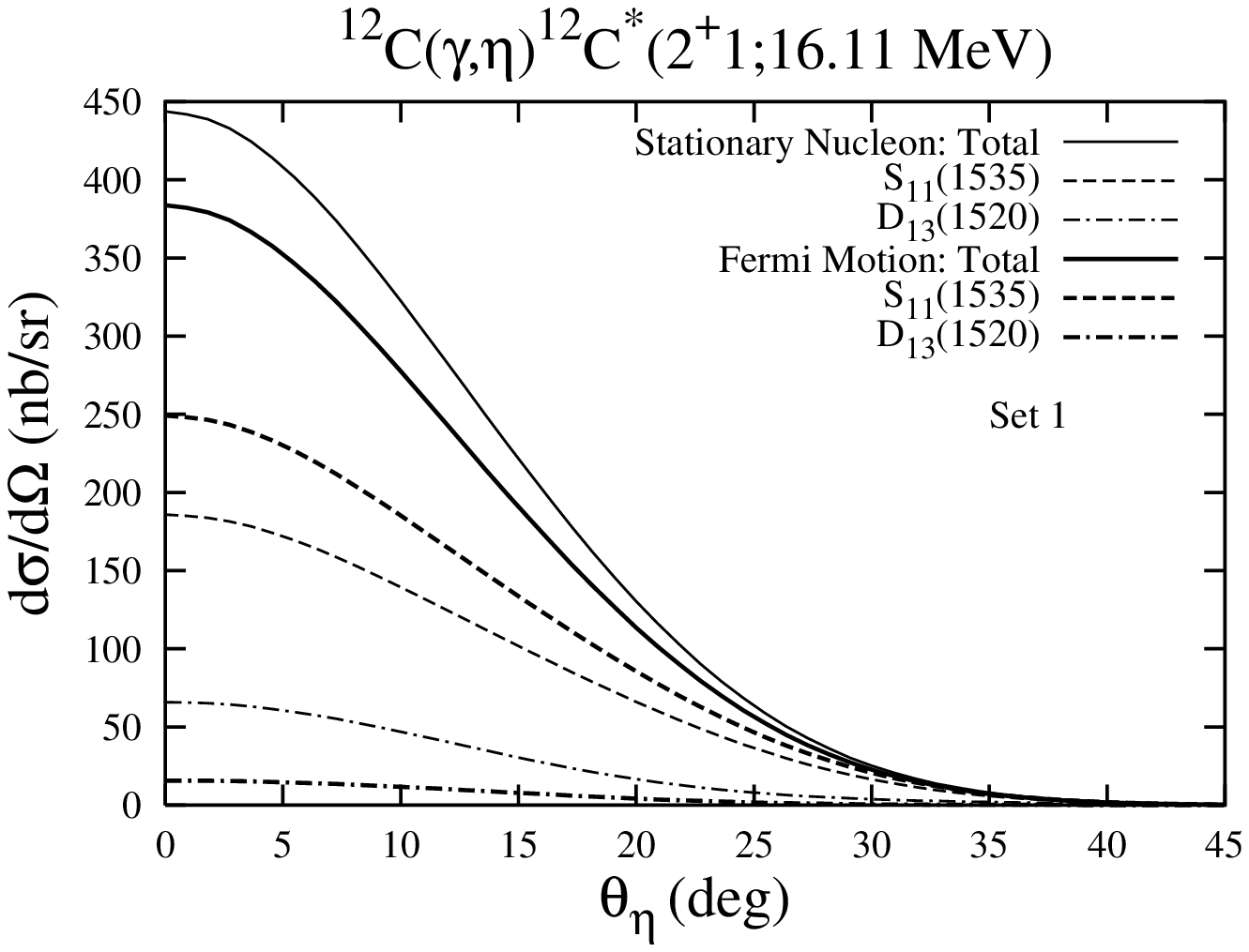}}
\end{center}

\newpage
\begin{center}
\resizebox{\textwidth}{!}{\includegraphics{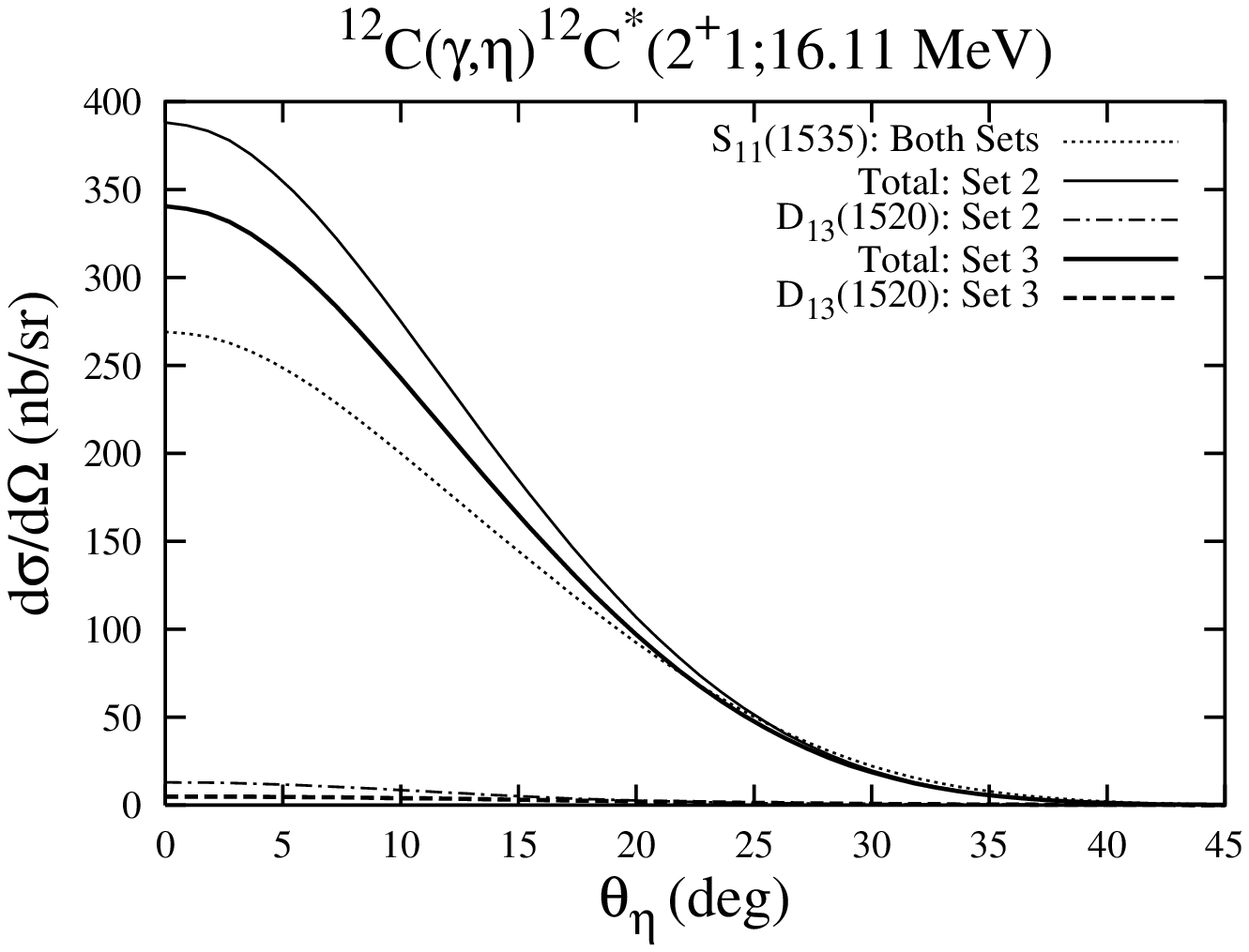}}
\end{center}

\newpage
\begin{center}
\resizebox{\textwidth}{!}{\includegraphics{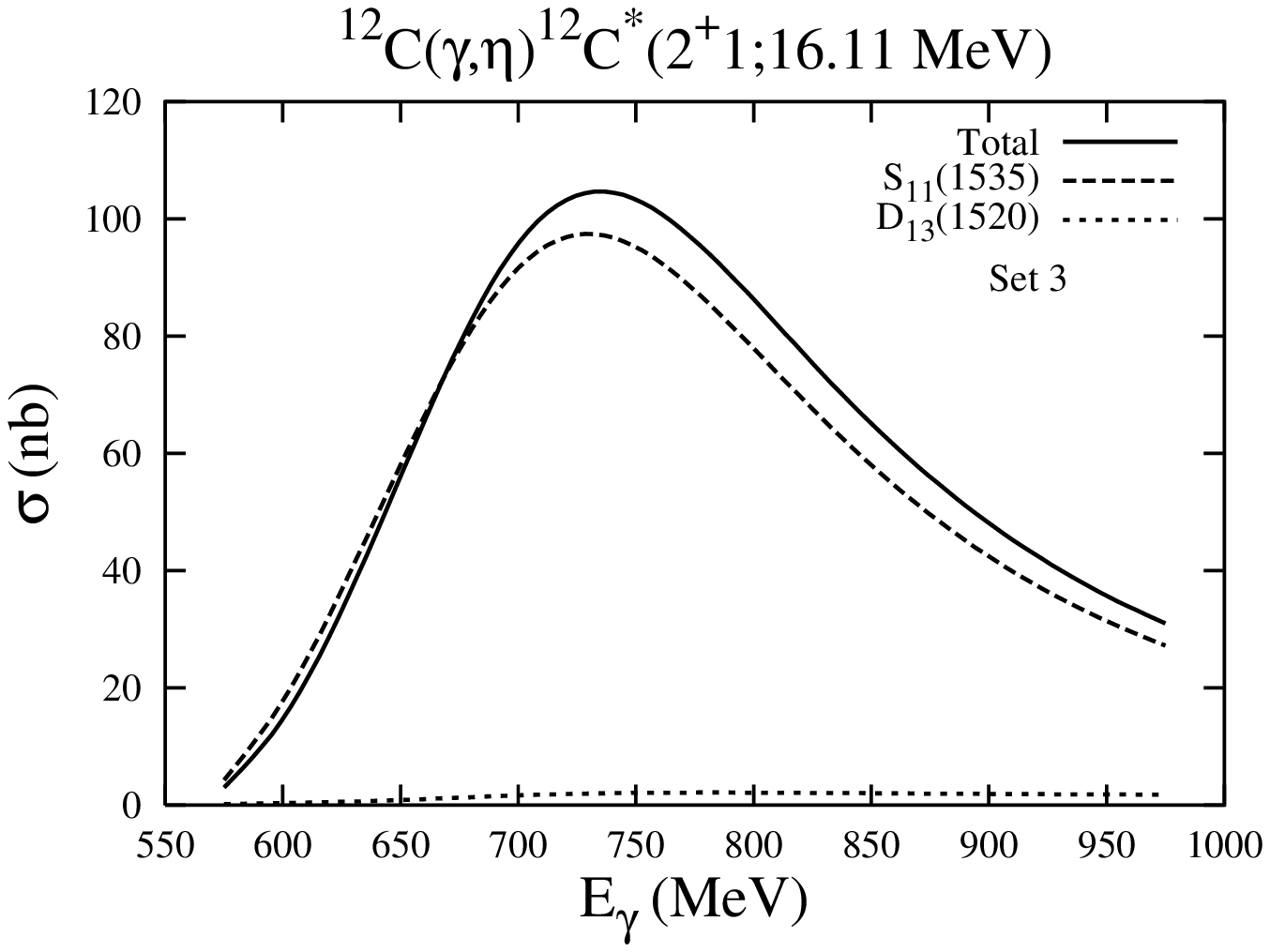}}
\end{center}

\newpage
\begin{center}
\resizebox{\textwidth}{!}{\includegraphics{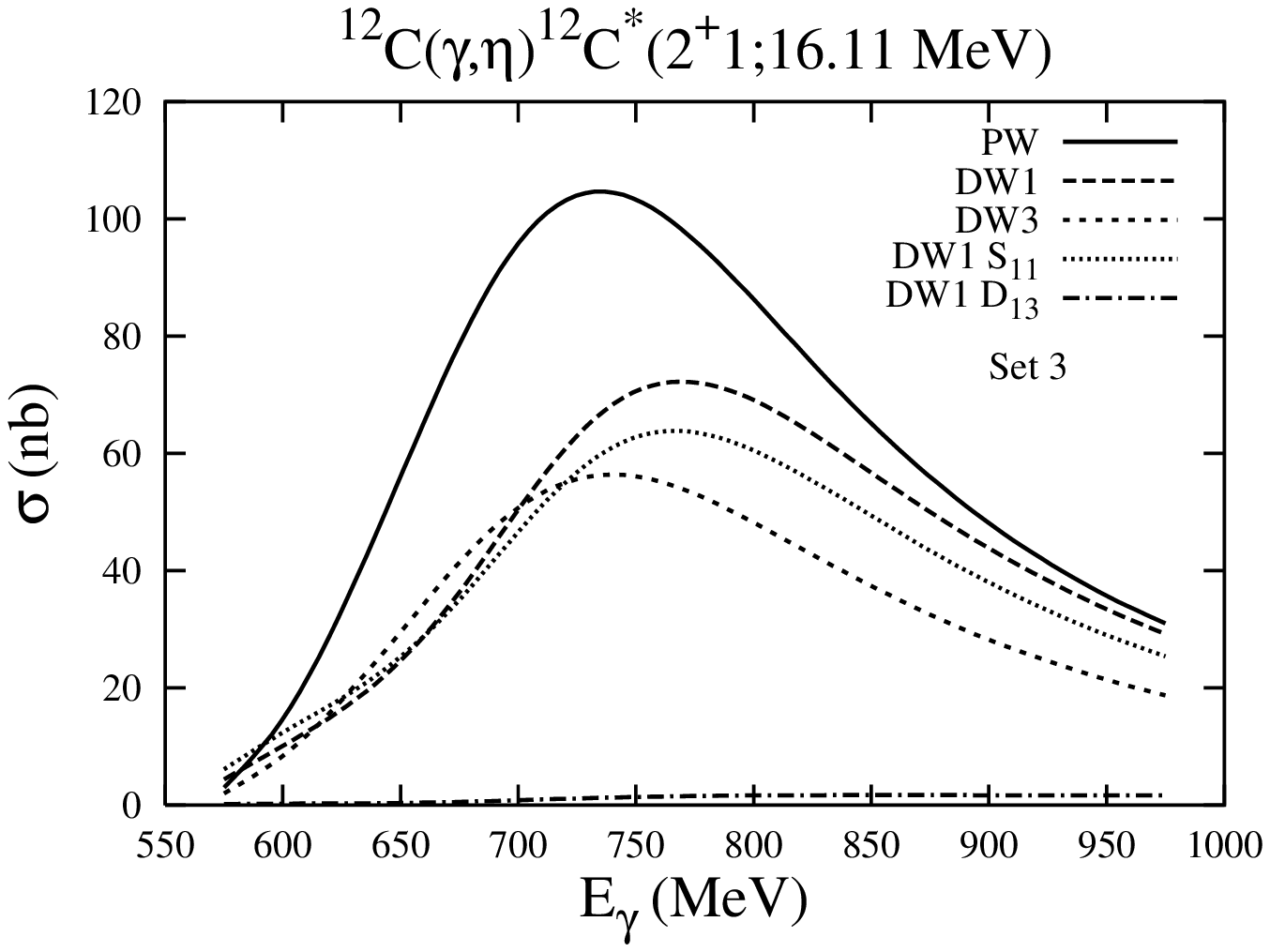}}
\end{center}

\newpage
\begin{center}
\resizebox{!}{\textheight}{\includegraphics{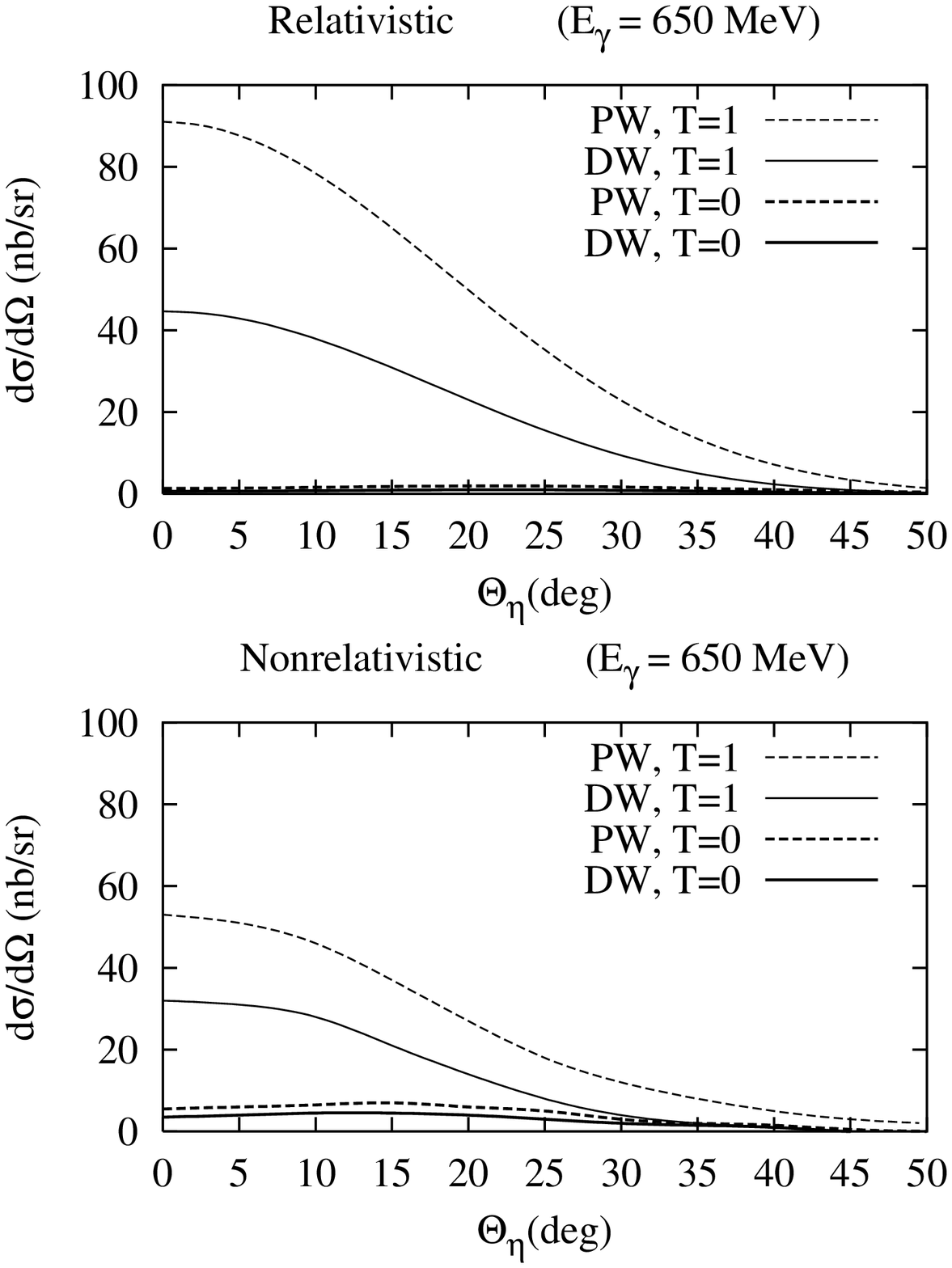}}
\end{center}

\newpage
\begin{center}
\resizebox{\textwidth}{!}{\includegraphics{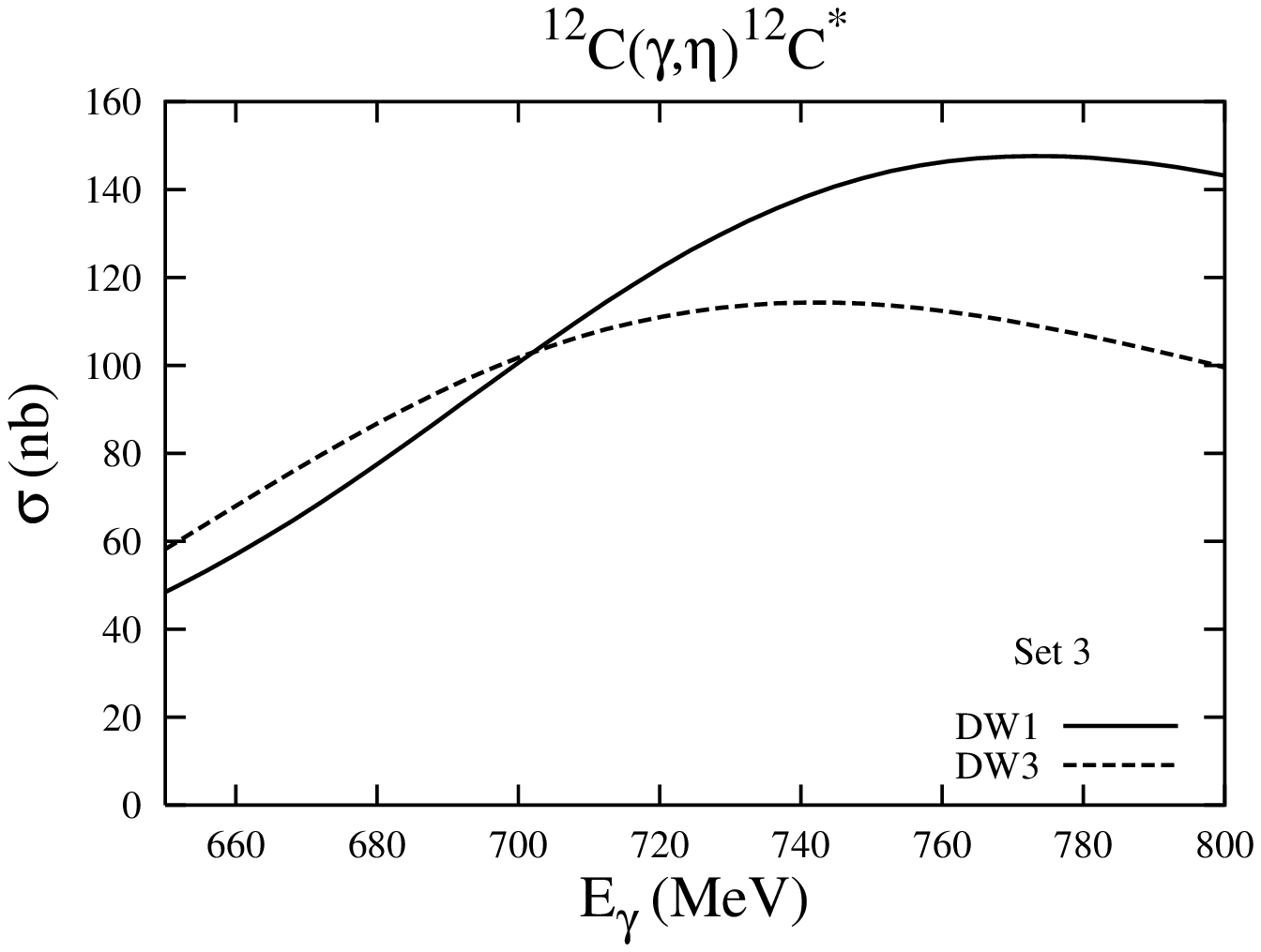}}
\end{center}

\end{document}